\documentstyle[12pt,a4,epsf]{article}
\begin{document}

\baselineskip=17pt plus 0.2pt minus 0.1pt


\makeatletter
\def\calN{{\cal N}}
\def\vF{{\mbox{\boldmath $F$}}}
\def\vB{{\mbox{\boldmath $B$}}}
\def\vp{{\mbox{\boldmath $\partial$}}}
\def\vx{{\mbox{\boldmath $x$}}}
\def\Tr{\mathop{\rm Tr}}
\def\tr{\mathop{\rm tr}}
\def\c{{\cos\phi}}
\def\s{{\sin\phi}}
\def\cPhi{{\bar\Phi}}
\def\cz{{\bar z}}
\def\cp{{\bar\partial}}
\newcommand{\vs}[1]{\vspace*{#1}}
\newcommand{\hs}[1]{\hspace*{#1}}
\newcommand{\PB}[2]{\{#1,#2\}}
\newcommand{\comm}[2]{\left[#1,#2\right]}
\def\p{{\partial}}
\def\nn{{\nonumber}}
\renewcommand{\thefootnote}{\fnsymbol{footnote}}
\newcommand{\bra}[1]{\left\langle #1\right|}
\newcommand{\ket}[1]{\left| #1\right\rangle}
\newcommand{\VEV}[1]{\left\langle #1\right\rangle}
\newcommand{\braket}[2]{\VEV{#1 | #2}}
\newcommand{\be}{\begin{equation}}
\newcommand{\ee}{\end{equation}}
\newcommand{\bea}{\begin{eqnarray}}
\newcommand{\eea}{\end{eqnarray}}
\newcommand{\diag}{\mathop{\rm diag}}
\newcommand{\Pf}{\mathop{\rm Pf}}
\begin{titlepage}
\title{
\hfill\parbox{4cm}
{\normalsize KUNS-1654\\{\tt hep-th/0003231}}\\
\vspace{1cm}
Noncommutative Monopole\\from Nonlinear Monopole
}
\author{
Sanefumi {\sc Moriyama}
\thanks{{\tt moriyama@gauge.scphys.kyoto-u.ac.jp}}
\\[7pt]
{\it Department of Physics, Kyoto University, Kyoto 606-8502, Japan}
}
\date{\normalsize March, 2000}
\maketitle
\thispagestyle{empty}

\begin{abstract}
\normalsize
We solve the non-linear monopole equation of the Born-Infeld theory to 
all orders in the NS 2-form and give physical implications of the
result. The solution is constructed by extending the earlier idea of
rotating the brane configuration of the Dirac monopole in the target
space. After establishing the non-linear monopole, we explore the
non-commutative monopole by the Seiberg-Witten map.
\end{abstract}

\end{titlepage}

\section{Introduction}
Recently non-commutative gauge theory has received much attention for
its origin in string theory.
The effective action of D-brane in string theory with a constant NS
2-form $B_{ij}$ is the non-commutative Born-Infeld theory when the
point splitting regularization is adopted \cite{CDS,DH}.
On the other hand, if we adopt the Pauli-Villars regularization we
obtain the ordinary Born-Infeld theory.
Since the method of regularization should not change the physical
S-matrices, the two descriptions are argued to be related by a field
redefinition \cite{SW} (called the Seiberg-Witten map).

This relation has been investigated intensively from various aspects,
and among other things, from various BPS solutions
\cite{HasHas,HasHatMor,Bak,HatMor,Mat,HasHir}.
Since the constant NS 2-form serves as a uniform magnetic field, if
we view the monopole solution as a D-string ending on a D3-brane, we
expect the D-string tilts due to force balance between the D-string
tension and the magnetic force at the endpoint \cite{HasHas}.
This system was analyzed in \cite{KHas} as the solution of the
linearly realized BPS equation in the commutative space.

However if we would like to see the tilts directly from the
non-commutative viewpoint, it would be a hard task. We can only solve
the linearly realized BPS equation in the perturbation expansion with
respect to the non-commutativity parameter $\theta$.
And even if we have solved the BPS equation, we would have to know how
to extract the eigenvalues for the brane interpretation of 
\cite{CM,AHas}.
In our previous works \cite{HasHatMor,HatMor} we proposed the
non-commutative eigenvalue equation to analyze the asymptotic behavior
and confirmed the tilts.

In \cite{Mat,HasHir}, it was claimed that the brane interpretation is
possible if we transform the results into the commutative viewpoint by
the Seiberg-Witten map.
Actually they argued that the linearly realized BPS monopole in the
non-commutative space is mapped to the non-linearly realized BPS
monopole in the commutative space by extending the argument in the
instanton case \cite{SW}.
This relation is persuasive for the following reason.
Since we are considering the monopole in the non-commutative space
with the property that the field strength and the covariant derivative 
of the Higgs field vanish at the infinity and this property seems
unchanged under the Seiberg-Witten map, it is expected that what
relates to the non-commutative monopole by the Seiberg-Witten map
should also have the same property.
The condition of preserving the combination of supersymmetries which
is unbroken at the infinity where the field strength and the covariant
derivative vanish is exactly the non-linearly realized BPS equation.

One might worry that the linear BPS equation in the non-commutative
space usually derived from the Yang-Mills theory cannot be
obtained from the Born-Infeld theory which is of our prime interest in 
discussing the Seiberg-Witten map.
However, it is shown that even in the non-commutative space the linear
BPS equation of the Yang-Mills theory reproduces the equation of
motion of the Born-Infeld theory by extending earlier discussions for
the non-Abelian Born-Infeld theory \cite{AHas} if we adopt the
symmetrized trace prescription \cite{STr} for the definition of the
determinant.
Note also that this kind of transformation is possible owing to
another fact that the solution is unchanged even when the derivative
corrections to the Born-Infeld action are taken into account
\cite{Tho}.

Moreover it is proposed that the non-linear BPS monopole in the
commutative space is related to the linear BPS monopole in the
commutative space by the rotation in the target space.
In the electric case of \cite{Mat}, an exact treatment of the soliton
was given though there are no discussions on the non-linear BPS
equation. In the magnetic case of \cite{HasHir} the discussion on the
non-linear BPS equation was restricted to the approximation
$r^2\gg 2\pi\alpha'\gg(2\pi\alpha')^2B$.

In this paper we shall extend the works of \cite{Mat,HasHir} to
explore the non-commutative BPS monopole from the non-linear BPS
monopole in the commutative space.
First we shall solve the non-linear BPS equation in the commutative
space exactly without any approximation.
We find that the solution is nothing but the one obtained by rotating
the solution of the linear BPS equation in the target space.

After establishing the non-linear BPS monopole in the commutative
space, we explicitly write down the first few terms in the expansion
of the NS 2-form $B_{ij}$.
What we find is terms in a mess and at first sight it seems impossible 
that they are related to the non-commutative monopole by the
Seiberg-Witten map.
We shall resolve this problem by using the moduli of the open string,
namely, the open string metric $G$ and the non-commutativity parameter
$\theta$. 
This resolution is regarded as an evidence for the claim that the
non-linear monopole is transformed into the non-commutative monopole
by the Seiberg-Witten map.
Finally we map the non-linear monopole into the non-commutative
space. We confirm that it satisfies the non-commutative BPS equation
up to $O(\theta^2)$.

\section{Nonlinear BPS equation}
In this section, we shall explicitly solve the non-linear BPS equation 
in the commutative space.
First we shall recall the linearly realized supersymmetries
$\delta_{\rm L}$ and non-linearly realized supersymmetries
$\delta_{\rm NL}$ \cite{BG,Ket,Tse,SW} of the Born-Infeld 
action:
\bea
\delta_{\rm L}\lambda
&=&\frac{1}{2\pi\alpha'}M_{mn}^+\sigma^{mn}\eta,\\
\delta_{\rm L}\bar\lambda
&=&\frac{1}{2\pi\alpha'}M_{mn}^-\sigma^{mn}\bar\eta,\\
\delta_{\rm NL}\lambda&=&\frac{1}{4\pi\alpha'}
\Bigl(1-\Pf M+\sqrt{1-\Tr M^2/2+(\Pf M)^2}\Bigr)\eta^*,\\
\delta_{\rm NL}\bar\lambda&=&\frac{1}{4\pi\alpha'}
\Bigl(1+\Pf M+\sqrt{1-\Tr M^2/2+(\Pf M)^2}\Bigr)\bar\eta^*,
\eea
where $M$ denotes
\bea
M=(2\pi\alpha')\pmatrix{0&-\p_1\Phi&-\p_2\Phi&-\p_3\Phi\cr
\p_1\Phi&0&(F_3+B_3)&-(F_2+B_2)\cr
\p_2\Phi&-(F_3+B_3)&0&(F_1+B_1)\cr
\p_3\Phi&(F_2+B_2)&-(F_1+B_1)&0},
\eea
with the magnetic field $F_i=\epsilon_{ijk}F_{jk}/2$, a constant NS
2-form background $B_i=\epsilon_{ijk}B_{jk}/2$ and the Higgs field
$\Phi$.
Here we turn on only the spatial components of the field strength and
the NS 2-form.
The matrix $M$ has been obtained by regarding the Euclidean time
component of the gauge field as the Higgs field $\Phi$ and discarding
the time derivatives.
We shall set $2\pi\alpha'=1$ for simplicity hereafter, however
we can restore it on the dimensional ground anytime we like.
The non-linear BPS equation \cite{SW,MMMS} is the condition for
preserving the linear combination of $\delta_{\rm L}$ and 
$\delta_{\rm NL}$ which is unbroken at the infinity where the field
strength and the derivative of the Higgs field vanish:
\bea
\frac{\vF+\vB-\vp\Phi}{1+(\vF+\vB)\cdot\vp\Phi+
\sqrt{1+(\vF+\vB)^2+(\vp\Phi)^2+\Bigl((\vF+\vB)\cdot\vp\Phi\Bigr)^2}}
=\frac{\vB}{1+\sqrt{1+\vB^2}}.
\label{BPS}
\eea

This BPS equation is not so complicated to solve as it looks.
The starting point is similar to the case of instanton \cite{Ter}.
First we note that eq.\ (\ref{BPS}) implies $\vF-\vp\Phi$ is
proportional to $\vB$:
\bea
\vF-\vp\Phi=f\vB,
\label{F-pPhi}
\eea
where $f$ is an unknown function.
The key point to solve this BPS equation is to rewrite eq. (\ref{BPS})
as
\bea
(f+1)\Bigl(1+\sqrt{1+\vB^2}\Bigr)-1-(\vF+\vB)\cdot\vp\Phi
=\sqrt{1+(\vF+\vB)^2+(\vp\Phi)^2+\Bigl((\vF+\vB)\cdot\vp\Phi\Bigr)^2}.
\label{square}
\eea
Taking the square of this equation (\ref{square}) and using the
relation (\ref{F-pPhi}) to eliminate the magnetic field $\vF$ when
necessary, we find that eq.\ (\ref{square}) is reduced simply to
\bea
f=(\vp\Phi)^2+(f+1)\vB\cdot\vp\Phi.
\label{f=Pf}
\eea
Another equation for $f$ and $\Phi$ besides (\ref{f=Pf}) is obtained
by taking the divergence of the relation (\ref{F-pPhi}) and using the
Bianchi identity $\vp\cdot\vF=0$,
\bea
-\vp^2\Phi=\vB\cdot\vp f.
\label{bianchi}
\eea
Now we have a system of differential equations (\ref{f=Pf}) and
(\ref{bianchi}) for the scalar quantities $f$ and $\Phi$.
After eliminating $f$ we find quite a non-linear equation for $\Phi$:
\bea
\vp^2\Phi\Bigl(1-\vB\cdot\vp\Phi\Bigr)^2
+2\vB\cdot\vp\vp\Phi\cdot\vp\Phi\Bigl(1-\vB\cdot\vp\Phi\Bigr)
+\vB\cdot\vp\vB\cdot\vp\Phi\Bigl(1+(\vp\Phi)^2\Bigr)=0.
\label{higgsBPS}
\eea
Hereafter we shall suppose the constant background $\vB$ is in the $z$ 
direction and rewrite the equation (\ref{higgsBPS}) in the cylindrical
coordinate $(\rho,\varphi,z)$ with $x=\rho\cos\varphi$ and
$y=\rho\sin\varphi$,
\bea
&&(\p_\rho^2\Phi+\p_\rho\Phi/\rho+\p_z^2\Phi)(1-B\p_z\Phi)^2
+2B(\p_\rho\p_z\Phi\p_\rho\Phi+\p_z^2\Phi\p_z\Phi)(1-B\p_z\Phi)\nn\\
&&\hs{3cm}+B^2\p_z^2\Phi(1+(\p_\rho\Phi)^2+(\p_z\Phi)^2)=0.
\label{nonlinear}
\eea

This differential equation looks impossible to solve. However, we can
apply the idea of \cite{Mat,HasHir} to find that the solution is
exactly the one obtained by rotating the solution of the linear BPS
equation in the target space by an angle $\phi$ with $\tan\phi=B$. 
This idea is convincing for the following reason.\footnote{
We are grateful to T.~Hirayama for a valuable discussion on this
point.
}
Originally the string theory has the $SO(1,9)$ Lorentz symmetry and 32
supersymmetries.
Taking the static gauge the Lorentz symmetry is broken into
$SO(1,3)\times SO(6)$ and half of the supersymmetries are broken.
The broken symmetries are realized non-linearly.
If we rotate the target space and still take the static gauge by
adopting a different worldsheet coordinate, we would find that
originally linearly realized symmetries correspond in general to some
combinations of the linear and the non-linear ones.
Therefore the linear BPS equation and the non-linear BPS equation
should be related by a target space rotation.

\begin{figure}[hbt]
\begin{center}
\leavevmode
\epsfxsize=60mm
\put(55,85){$\Phi$}
\put(90,60){$z$}
\put(130,70){$\cPhi$}
\put(140,50){$\cz$}
\epsfbox{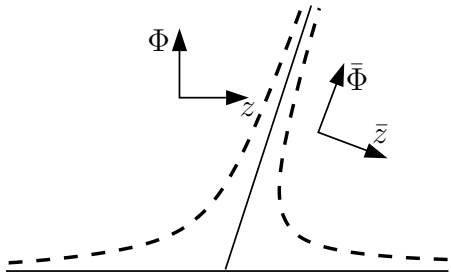}
\caption{It is easier to find the solution using the coordinate
  $(\cPhi,\cz)$ instead of $(\Phi,z)$. The dashed line denotes the
  solution of the Higgs field to be found.}
\label{nlBPS}
\end{center}
\end{figure}

To see this explicitly we change our variables into those with bars by
the target space rotation,
\bea
\pmatrix{\cPhi\cr\cz}
=\pmatrix{\cos\phi&\sin\phi\cr-\sin\phi&\cos\phi}\pmatrix{\Phi\cr z},
\label{rotation}
\eea
and show that the solution for the variables with bars is the same as
that of the linear BPS equation.
First we have to rewrite the equation (\ref{nonlinear}) by changing
$\Phi$, $\p_\rho$, $\p_z$ into $\cPhi$, $\cp_\rho\equiv\p/\p \rho|_\cz$,
$\cp_z\equiv\p/\p\cz|_\rho$.
Note that, though we do not change the coordinate $\rho$, $\cp_\rho$ is
different form $\p_\rho$ because the coordinate to be fixed is different
between them.
The formulas for rewriting $\p_z\Phi$ and $\p_\rho\Phi$ into $\cp_z\cPhi$ 
and $\cp_\rho\cPhi$ read
\bea
\p_z\Phi&=&\frac{\c\cp_z\cPhi-\s}{\c+\s\cp_z\cPhi},\\
\p_\rho\Phi&=&\frac{\cp_\rho\cPhi}{\c+\s\cp_z\cPhi},
\eea
where the first formula is directly obtained from the rotation
(\ref{rotation}) and the second formula is a consequence of the chain
rule formula 
\bea
\left.\frac{\p\cPhi}{\p \rho}\right|_{\mbox{$z$}}
=\left.\frac{\p \rho}{\p \rho}\right|_{\mbox{$z$}}
\left.\frac{\p\cPhi}{\p \rho}\right|_{\mbox{$\cz$}}
+\left.\frac{\p\cz}{\p \rho}\right|_{\mbox{$z$}}
\left.\frac{\p\cPhi}{\p\cz}\right|_{\mbox{$\rho$}},
\eea
and the relations $\p\cPhi/\p \rho|_z=\c\,\p\Phi/\p \rho|_z$ and 
$\p\cz/\p \rho|_z=-\s\,\p\Phi/\p \rho|_z$.
In the same way we also find the similar formulas for higher
derivatives,
\bea
\p_z^2\Phi&=&\cp_z^2\cPhi/(\c+\s\cp_z\cPhi)^3,\\
\p_z\p_\rho\Phi&=&\Bigl[\c\cp_z\cp_\rho\cPhi
+\s(\cp_z\cp_\rho\cPhi\cp_z\cPhi-\cp_z^2\cPhi\cp_\rho\cPhi)\Bigr]/
(\c+\s\cp_z\cPhi)^3,\\
\p_\rho^2\Phi&=&\Bigl[(\c)^2\cp_\rho^2\cPhi
+2\c\s\Bigl(-\cp_z\cp_\rho\cPhi\cp_\rho\cPhi
+\cp_\rho^2\cPhi\cp_z\cPhi\Bigr)\nn\\
&&+(\s)^2\Bigl(\cp_z^2\cPhi(\cp_\rho\cPhi)^2
-2\cp_z\cp_\rho\cPhi\cp_z\cPhi\cp_\rho\cPhi
+\cp_\rho^2\cPhi(\cp_z\cPhi)^2\Bigr)\Bigr]/(\c+\s\cp_z\cPhi)^3.
\eea
Using these formulas, the terribly non-linear equation
(\ref{nonlinear}) now becomes
\bea
\cp_\rho^2\cPhi+\cp_\rho\cPhi/\rho+\cp_z^2\cPhi=0,
\label{laplace}
\eea
which is nothing but the three-dimensional laplace equation.
The solution to eq.\ (\ref{laplace}) is given by the sum of the
Coulomb term and the linear term determined by the boundary
condition in the asymptotic region:
\bea
\cPhi=\frac{q}{\sqrt{\rho^2+\cz^2}}+B\cz.
\label{bar}
\eea
Turning back to the variables without bars using the relation
(\ref{rotation}), our final result for the Higgs
field $\Phi$ is given as the solution of the algebraic equation,
\bea
\Bigl((1+B^2)\rho^2+z^2-2Bz\Phi+B^2\Phi^2\Bigr)\Phi^2=q^2,
\label{result}
\eea
or its covariant form
\bea
\Bigl((1+\vB^2)\vx^2-(\vB\cdot\vx)^2
-2\vB\cdot\vx\Phi+\vB^2\Phi^2\Bigr)\Phi^2=q^2.
\label{covariant}
\eea
The explicit expression of the first few terms in the expansion with
respect to $B$ is 
\bea
\Phi=\frac{q}{r}
+\frac{q^2\vB\cdot\vx}{r^4}
-\frac12\frac{q\vB^2}{r}
+\frac12\frac{q(\vB\cdot\vx)^2}{r^3}
-\frac12\frac{q^3\vB^2}{r^5}
+\frac52\frac{q^3(\vB\cdot\vx)^2}{r^7},
\label{expand}
\eea
with $r=\sqrt{\rho^2+z^2}$.

Similarly the magnetic field is also obtained from the relations
(\ref{F-pPhi}) and (\ref{f=Pf}) as
\bea
\vF=\vp\Phi+\frac{(\vp\Phi)^2+\vB\cdot\vp\Phi}{1-\vB\cdot\vp\Phi}\vB.
\label{magnetic}
\eea
Using our result (\ref{covariant}) we can rewrite this expression 
(\ref{magnetic}) by eliminating the derivatives of the Higgs field
$\vp\Phi$,
\bea
\vF=\frac{(1+\vB^2)(-\vx+2\Phi\vB)\Phi}
{(1+\vB^2)\vx^2-(\vB\cdot\vx)^2-3(\vB\cdot\vx)\Phi+2\vB^2\Phi^2}.
\label{magcov}
\eea

\section{Physical interpretation}
\begin{figure}[hbt]
\begin{center}
\leavevmode
\epsfxsize=120mm
\put(80,90){$\Phi$}
\put(150,5){$z$}
\put(70,0){$\rho$}
\put(255,90){$\Phi$}
\put(335,5){$z$}
\put(80,-15){(A)}
\put(260,-15){(B)}
\put(100,100){D-string}
\put(140,30){D3-brane}
\epsfbox{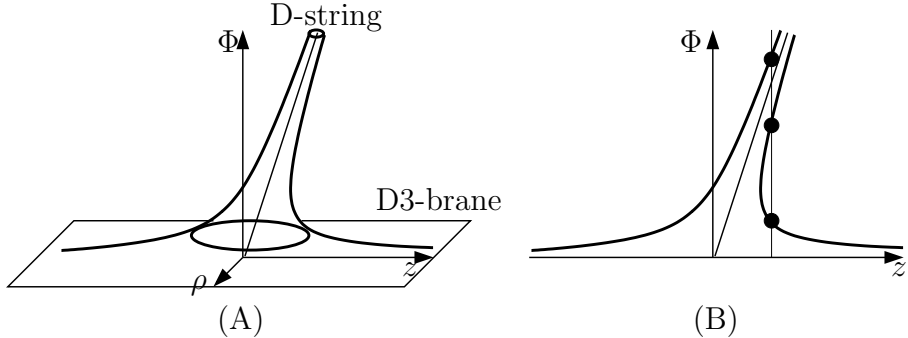}
\caption{The Higgs field of the Dirac monopole is depicted against the
  worldsheet coordinate $(z,\rho)$ in the left figure (A).
  The right  figure (B) is the one restricted to the $\rho=0$ plane.
  As seen from (B) the Higgs field is multi-valued for a sufficiently
  large $z$.}
\label{plot}
\end{center}
\end{figure}

In this section, we shall give some comments and physical
interpretations to our solution.
First, our result is obtained without any approximations and
the expansion (\ref{expand}) is consistent with the result
obtained in \cite{HasHir}.
The behavior of the Higgs field $\Phi$ (\ref{result}) against the
worldsheet coordinate $(z,\rho)$ is depicted in Fig.\ \ref{plot} (A).
Note that in the right hand side of eq. (\ref{square}) we do not
persist in taking the positive branch of the square root, because it
forces us to discard part of the solution given in Fig.\ \ref{plot}
(A).
The spike-like behavior of the Higgs field represents the D-string
attached to the D3-brane in the brane interpretation of
\cite{CM,AHas}.
This D-string tilts due to the uniform magnetic field and the tilt
angle is exactly the one expected from the force balance
\cite{HasHas}.
Here we find that the Higgs field is multi-valued due to this tilt
(see Fig.\ \ref{plot} (B) which shows the multi-valuedness of $\Phi$
on the $\rho=0$ plane).
This multi-valuedness is a consequence of the fact that the eq.\
(\ref{result}) determining $\Phi$ is a fourth order algebraic
equation which in general has four solutions.
Another solution not depicted in Fig.\ \ref{plot} (B) is a fake one
with $\Phi< 0$.
This multi-valuedness implies that the Dirac monopole might be
ill-defined as a field theoretic soliton in the non-linear BPS
equation and probably also in the non-commutative BPS equation via
the Seiberg-Witten map. However, the multi-valuedness is inevitable from
the string theory viewpoint.

\begin{figure}[hbt]
\begin{center}
\leavevmode
\epsfxsize=50mm
\put(65,20){$\vec v$}
\put(75,5){$\vec w$}
\epsfbox{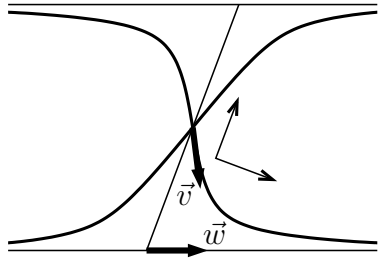}
\caption{The 't Hooft-Polyakov monopole is not multi-valued when the
  single-valued condition (\ref{single}) holds.}
\label{multi}
\end{center}
\end{figure}

Though we do not know the non-linear BPS equation for the non-Abelian
case due to the complexity of the ordering in the determinant, it is
expected that the Higgs field related to the non-commutative monopole by
the Seiberg-Witten map is that obtained by rotating the solution of
the linear BPS equation in the target space.
Note that in this case of the 't Hooft-Polyakov monopole, the
problematic multi-valuedness in the Dirac monopole does not
necessarily appear.
From the behavior near the origin $r=0$ of the exact solution in
\cite{Bog,PS} with $C=\langle\Phi\rangle$,
\bea
\Phi=(Cr/\tanh Cr-1)/r\sim C^2r/3,
\eea
we can read off that the tangent vector of the deformed D3-brane is
$\vec v=(1,-C^2/3+B)$ and that of the worldsheet parameterization is
$\vec w=(1,B)$ 
in the rotated coordinate system depicted in Fig.\ \ref{multi}.
Therefore the single-valuedness condition is expressed as the
positivity of the inner product of these two vectors:
\bea
\vec v\cdot\vec w=1-C^2B/3+B^2>0.
\label{single}
\eea
This implies that at some value of NS 2-form even the 't
Hooft-Polyakov monopole is not single-valued, which is something we
have never experienced in the usual field-theoretical solitons.

Finally, we would like to comment on the small $B$ expansion
(\ref{expand}). Unlike our experience of the non-commutative monopole 
in the flat space \cite{HasHatMor,HatMor} where the only parameter is
$\theta$, at $O(B^2)$ of (\ref{expand}) we find terms proportional to
$r^{-1}$ as well as $r^{-5}$, which implies the parameter
$2\pi\alpha'$ also appears.
If we expect the present result is transformed to the non-commutative
monopole by the Seiberg-Witten map, this kind of double expansion seems
impossible.

The resolution to this paradox is given by considering the moduli of
the open string metric $G$ and the non-commutativity parameter
$\theta$.
When we relate the non-commutative gauge theory to its commutative
counterpart, we should also relate the moduli by \cite{SW}
\bea
\frac{1}{G}+\theta=\frac{1}{g+B}.
\eea
Since we set the metric in the commutative space to the flat one
$g_{ij}=\delta_{ij}$ and turn on only the spatial NS 2-form $B_i$, our 
moduli are
\bea
G_{ij}=(1+\vB^2)\delta_{ij}-B_iB_j,
\qquad
\theta^{ij}=-\frac{\epsilon_{ijk}B_k}{1+\vB^2},
\eea
with a necessarily non-trivial open string metric.
Using these open string moduli we can construct several kinds of
scalars:
\bea
R^2&\equiv&G_{ij}x^ix^j=(1+B^2)\rho^2+z^2,\\
\theta\cdot x&\equiv&\sqrt{G}\epsilon_{ijk}\theta^{jk}x^i=-Bz,\\
\theta^2&\equiv&G^{ij}\sqrt{G}\epsilon_{ikl}\theta^{kl}
\sqrt{G}\epsilon_{jmn}\theta^{mn}=B^2.
\eea
Note that in the case of a non-trivial metric the $\epsilon$ tensor
should always be accompanied with $\sqrt{G}$.
In terms of these scalars our result (\ref{result}) can be rewritten
into an expression with only one parameter $\theta$:
\bea
\Bigl(R^2+2\theta\cdot x\Phi+\theta^2\Phi^2\Bigr)\Phi^2=q^2.
\label{modchange}
\eea
Note that from the dimensional ground there can appear no
$2\pi\alpha'$ in (\ref{modchange}).
Our observation here shows that we have two viewpoints for the
non-linear monopole.
One is with the flat space and a NS 2-form and the other is with the
non-trivial metric and the non-commutativity parameter.
Similarly if we rewrite the magnetic field (\ref{magcov}) into the
covariant field strength, we will also find an expression without
$2\pi\alpha'$:
\bea
F_{ij}=\frac{\sqrt{G}\epsilon_{ijk}(-x^k-2\Phi\theta^k)\Phi}
{R^2+3\theta\cdot x\Phi+2\theta^2\Phi^2}.
\label{fieldstrength}
\eea
Similar expression for the gauge field is difficult to find because of
the existence of the Dirac string.

Our analysis so far is believed to be related to the non-commutative
monopole by the Seiberg-Witten map \cite{SW,Mat,HasHir}.
In the case of a constant non-trivial metric the BPS equation in the
non-commutative space should be given by
\bea
\widehat F_{ij}=\sqrt{G}\epsilon_{ijm}G^{mn}\widehat{D}_n\widehat\Phi,
\eea
as can be seen from the BPS bound arguments \cite{Bog,PS}.
However since the non-trivial metric is constant, we can always
orthonormalize it globally by the vielbein:
\bea
E_\alpha^iE_\beta^jG_{ij}=\delta_{\alpha\beta}.
\eea
Therefore if we would like to find the non-commutative monopole in the 
flat space we have to collect all our results of the non-linear BPS
equation, rewrite them in the covariant form, make a coordinate
transformation $x^i\to E_i^\alpha x^i$ into the flat space, and
transform them into the non-commutative space by the Seiberg-Witten
map.
Our result in the flat space up to $O(\theta^2)$ for the Higgs field
is
\bea
\Phi=\frac{q}{r}-\frac{q^2\theta\cdot x}{r^4}
-\frac{q^3\theta^2}{2r^5}+\frac{5q^3(\theta\cdot x)^2}{2r^7},
\eea
where we have rewritten $R$ into $r$ because now we are in the flat
space.
And the gauge field corresponding to the field strength
(\ref{fieldstrength}) is given by
\bea
A_i=A_i^0+A_i^1,
\label{gaugesol01}
\eea
with
\bea
&&A_1^0=\frac{qy}{r(r+z)},\quad A_2^0=-\frac{qx}{r(r+z)},\quad
A_3^0=0,
\label{gaugesol0}\\
&&A_i^1=\frac{q^2\epsilon_{ijk}\theta_jx_k}{r^4}
-\frac{5q^3\epsilon_{ijk}\theta_jx_k\theta_mx_m}{2r^7}.
\label{gaugesol1}
\eea
Note that due to the presence of the Dirac string the solution in the
zero-th order in $\theta$ cannot be written in a spherically symmetric
form.
We have explicitly transformed this result into the non-commutative
space by the Seiberg-Witten map \cite{Oku}
\bea
\widehat A_i&=&A_i-\frac12\theta^{kl}A_k(\p_lA_i+F_{li})
+\frac12\theta^{kl}\theta^{mn}
A_k(\p_lA_m\p_nA_i-\p_lF_{mi}A_n+F_{lm}F_{ni}),\\
\widehat\Phi&=&\Phi-\frac12\theta^{kl}A_k(2\p_l\Phi)
+\frac12\theta^{kl}\theta^{mn}
A_k(\p_lA_m\p_n\Phi-\p_l\p_m\Phi A_n+F_{lm}\p_n\Phi),
\eea
and checked that it indeed satisfies the non-commutative BPS equation
by using a symbolic manipulation software.
However due to the Dirac string the covariant form is not available
and the result is too complicated and not suitable to be written here.

\section{Summary and further directions}
In this paper we extended the earlier idea of rotating the system to
solve the non-linear BPS equation without any approximation.
Since we solved it exactly, the multi-valuedness problem appeared.
We also pointed out the open string metric is in general non-trivial
and a careful treatment is necessary.
Finally we transformed our result into the non-commutative space by
the Seiberg-Witten map and confirmed it satisfies the non-commutative
BPS equation.

In our exact manipulation we clarified the physical meaning of
the Higgs field in the non-linear BPS equation.
Hence in the non-Abelian case, even though we do not know the
non-linear BPS equation, we expect the solution for the Higgs field
related to the non-commutative monopole is that obtained by rotating
the solution of the linear BPS equation in the target space.
However the meaning of the gauge field is still unclear.
So we do not know what to expect for the gauge field.
To understand it is an interesting subject.

{\bf Acknowledgment}

We would like to thank K.\ Hashimoto, H.\ Hata and T.\ Hirayama for
valuable discussions and comments and H.\ Hata for careful reading of
the present manuscript.
This work is supported in part by Grant-in-Aid for Scientific Research
from Ministry of Education, Science, Sports and Culture of Japan
(\#04633).
The author is supported in part by the Japan Society for the
Promotion of Science under the Predoctoral Research Program.

\newcommand{\J}[4]{{\sl #1} {\bf #2} (#3) #4}
\newcommand{\andJ}[3]{{\bf #1} (#2) #3}
\newcommand{\AP}{Ann.\ Phys.\ (N.Y.)}
\newcommand{\MPL}{Mod.\ Phys.\ Lett.}
\newcommand{\NP}{Nucl.\ Phys.}
\newcommand{\PL}{Phys.\ Lett.}
\newcommand{\PR}{ Phys.\ Rev.}
\newcommand{\PRL}{Phys.\ Rev.\ Lett.}
\newcommand{\PTP}{Prog.\ Theor.\ Phys.}
\newcommand{\hep}[1]{{\tt hep-th/{#1}}}


\begin{thebibliography}{99}
\bibitem{CDS}
A.~Connes, M.~R. Douglas and A.~Schwarz, ``Noncommutative Geometry
and Matrix Theory: Compactification on Tori'',
{\em JHEP} {\bf 02} (1998) 003,
{{\tt hep-th/9711162}}.

\bibitem{DH}
M.~R. Douglas and C.~Hull, ``D-branes and the Noncommutative Torus'',
{\em JHEP} {\bf 02} (1998) 008,
{{\tt hep-th/9711165}}.

\bibitem{SW}
N.~Seiberg and E.~Witten,
``String Theory and Noncommutative Geometry'',
{\em JHEP} {\bf 09} (1999) 032,
{{\tt hep-th/9908142}}.

\bibitem{HasHas}
A.~Hashimoto and K.~Hashimoto,
``Monopoles and Dyons in Non-Commutative Geometry'',
{\em JHEP} {\bf 11} (1999) 005,
{{\tt hep-th/9909202}}.

\bibitem{HasHatMor}
K.~Hashimoto, H.~Hata and S.~Moriyama, ``Brane Configuration from
Monopole Solution in Non-Commutative Super Yang-Mills Theory'',
{\em JHEP} {\bf 12} (1999) 021,
{{\tt hep-th/9910196}}.

\bibitem{Bak}
D.~Bak, ``Deformed Nahm Equation and a Noncommutative BPS Monopole'',
{\em Phys. Lett.} {\bf B471} (1999) 149, {{\tt hep-th/9910135}}.

\bibitem{HatMor}
H.~Hata and S.~Moriyama,
``String Junction from Non-Commutative Super Yang-Mills Theory'',
{\em JHEP} {\bf 0002} (2000) 011, {{\tt hep-th/0001135}}.

\bibitem{Mat}
D.~Mateos,
``Non-commutative vs. Commutative Descriptions of D-brane BIons'',
{{\tt hep-th/0002020}}.

\bibitem{HasHir}
K.~Hashimoto and T.~Hirayama, ``Branes and BPS Configurations of
Non-Commutative/ Commutative Gauge Theories'',
{{\tt hep-th/0002090}}

\bibitem{KHas}
K.~Hashimoto, ``Born-Infeld Dynamics in Uniform Electric Field'',
{\em JHEP} {\bf 07} (1999) 016,
{{\tt hep-th/9905162}}.

\bibitem{CM}
C.~G.~Callan Jr.~and J.~M.~Maldacena,
``Brane Dynamics From the Born-Infeld Action'',
{\em Nucl. Phys.} {\bf B513} (1998) 198,
{{\tt hep-th/9708147}}.

\bibitem{AHas}
A.~Hashimoto, ``The Shape of Branes Pulled by Strings'',
{\em Phys. Rev.} {\bf D57} (1998) 6441,
{{\tt hep-th/9711097}}.

\bibitem{STr}
A.~A.~Tseytlin,
``On non-abelian generalisation of Born-Infeld action in string theory'',
\J{\NP}{B501}{1997}{41}, {\tt hep-th/9701125}.

\bibitem{Tho}
L.~Thorlacius,
``Born-Infeld String as a Boundary Conformal Field Theory'',
\J{\PRL}{80}{1998}{1588}, {\tt hep-th/9710181}.

\bibitem{BG}
J.~Bagger and A.~Galperin, 
``New Goldstone multiplet for partially broken supersymmetry'', 
\J{\PR}{D55}{1997}{1091}, {\tt hep-th/9608177}.

\bibitem{Ket}
S.~V.~Ketov,
``A manifestly $N=2$ supersymmetric Born-Infeld action'',
\J{Mod.\ Phys.\ Lett.}{A14}{1999}{501}, {\tt hep-th/9809121}.

\bibitem{Tse}
A.~A.~Tseytlin,
``Born-Infeld action, supersymmetry and string theory'',
{\tt hep-th/9908105}.

\bibitem{MMMS}
M.~Marino, R.~Minasian, G.~Moore and A.~Strominger,
``Nonlinear Instantons from Supersymmetric $p$-Branes'',
\J{JHEP}{0001}{2000}{005}, {\tt hep-th/9911206}.

\bibitem{Ter}
S.~Terashima, ``$U(1)$ Instanton in Born-Infeld Action and
Noncommutative Gauge Theory'', {\tt hep-th/9911245}.

\bibitem{Bog}
E.~B.~Bogomol'nyi, ``The Stability of Classical Solutions'',
{\em Sov. J. Nucl. Phys.} {\bf 24} (1976) 449.

\bibitem{PS}
M.~K.~Prasad and C.~M.~Sommerfield, ``An Exact Classical Solution for
the 't Hooft Monopole and the Julia-Zee Dyon'',
{\em Phys. Rev. Lett.} {\bf 35} (1975) 760.

\bibitem{Oku}
K.~Okuyama, 
``A Path Integral Representation of the Map 
between Commutative and Noncommutative Gauge Fields'',
{\tt hep-th/9910138}.

\end{thebibliography}
\end{document}